\documentclass[twocolumn,amsmath,prl]{revtex4}

\usepackage{dcolumn}% Align table columns on decimal point
\usepackage{bm}% bold math
\usepackage[T1]{fontenc}
\usepackage{amsmath}
\usepackage{graphicx}

\begin{document}
\title{Engineering multiferroism in CaMnO$_3$}

\author{Satadeep Bhattacharjee, Eric Bousquet and Philippe Ghosez}
\affiliation{Physique Th\'eorique des Mat\'eriaux, Universit\'e de Li\`ege (B5), B-4000 Sart Tilman, Belgium}

\begin{abstract}
From first-principles calculations, we investigate the structural instabilities of CaMnO$_3$. We point out that, on top of a strong antiferrodistortive instability responsible for its orthorhombic ground-state, the cubic perovskite structure of CaMnO$_3$ also exhibit a weak ferroelectric instability.  Although ferroelectricity is suppressed by antiferrodistortive oxygen motions, we show that it can be favored using strain or chemical engineering in order to make CaMnO$_3$ multiferroic. We finally highlight that the FE instability of CaMnO$_3$ is Mn-dominated. This illustrates that, contrary to the common believe, ferroelectricity and magnetism are not necessarily exclusive but can be driven by the same cation.
\end{abstract}

 \maketitle

The family of ABO$_3$ perovskite oxide compounds consitutes an important class of multifunctional materials which, within the same simple cubic structure, exhibit a wide variety of behaviors and range from ferroelectric, piezoelectric or non-polar anti-ferrodistorted insulators to metals or superconductors, eventually combining an additional magnetic order \cite{book}. Recently, the search for magneto-electric multiferroics  operating at room temperature has motivated the study of oxides combining ferroelectric and magnetic orders like BiFeO$_3$ or YMnO$_3$ \cite{rev1,rev2}. In this framework, the case of CaMnO$_3$, a magnetic insulator which remains paraelectric, was considered as a prototypical example to reveal the antagonist role of partial $d$-shell occupacy in respectively favoring magnetism but penalizing ferroelectricity\cite{fil}.

CaMnO$_3$ is a G-type antiferromagnetic insulator\cite{mat}. It cristallizes at low temperature in a paraelectric \textit{Pnma}  orthorhombic ground-state structure which can be viewed as a small distortion of the cubic perovskite structure, produced by the tilting of oxygen octahedra (a$^-$b$^+$a$^-$)\cite{poepp,ab02}. Because of the Mn$^{4+}$ configuration, CaMnO$_3$ does not exhibit Jahn-Teller distortion : in the cubic structure,  the 5$d$ orbitals of Mn ion split into 3-fold degenerated t$_{2g}$ levels and two-fold degenerated e$_g$ levels with occupation being t$^3_{2g}$ and e$^0_g$\cite{k0}. CaMnO$_3$ does not either exhibit ferroelectric distortion.  In ABO$_3$ compounds, ferroelectricty is usually related to the O 2p -- B d hybridization \cite{Cohen} and Fillipetti and Hill \cite{fil} argued that because of the partial d-shell occupancy of Mn in CaMnO$_3$, Mn-ion is unable to accept a transfer of charge from the neighboring oxygen so that there is a strong resistance of the filled d-shells with spherical symmetry to move off-center. The fact that ferroelectricity requires d$^0$-ness while magnetism requires partial d-state occupancy is often evoked to explain the scarcity of compounds combining both properties \cite{fil,why,rev1,rev2}. It also suggests that the search for multiferroic ABO$_3$ perovskites should be oriented toward materials in  which magnetism and ferroelectricity are driven independently by the A and B cations \cite{rev2,Fi}. 

Khomskii {\it et. al} \cite{khom} further argued that since Mn ion is in the Mn$^{4+}$ state, the empty e$_g$ orbitals could nevertheless take part in the charge transfer process.  However, the authors suggested  that the exactly half-filled t$_{2g}$ orbitals might restrict the bonding between empty e$_g$ orbitals and oxygen p-orbitals by imposing Hund's rule on the e$_g$ states. More recently, a first-principles study of the othorhombic phase of CaMnO$_3$ \cite{sat} revealed that the static dielectric constant in that phase is very large and comparable to that of isostructural CaTiO$_3$ \cite{orthoCTO}. Since the latter is an incipient ferroelectric and its large dielectric response is coming from the softening of the highly polar ferroelectric mode, this is questioning the eventual tendency of CaMnO$_3$ to similarly exhibit tendency to ferroelectricity.

In this letter, we first carefully reinvestigate from first-principles the structural instabilities of cubic CaMnO$_3$. We point out that, contrary to common believe, CaMnO$_3$ do develop a weak ferroelectric (FE) instability at its equilibrium volume. However, like in CaTiO$_3$, the latter is hidden by a much deeper antiferrodistortive (AFD) instability which is responsible for the orthorhombic ground-state.  FE and AFD instabilities are however in competition in CaMnO$_3$ and we then explore the possibility of making it multiferroic using either strain or chemical engineering.

Our first-principles calculations have been performed in the framework of  Density Functional Theory (DFT) using the ABINIT package~\cite{abinit}.  In all our calculations we imposed a G-type antiferromagnetic order and worked within the Generalized Gradient Approximation (GGA), using the functional recently proposed by Wu and Cohen~\cite{wu2006}, which is known to be very accurate in predicting the unit-cell volume of solids. This choice is particularly important in the present study since ferroelectricity is strongly sensitive to the volume and accurate volume estimate is mandatory to make trustable predictions. We used optimized pseudopotentials\cite{rrkj} generated with OPIUM code\cite{opi}, treating 3$s$,3$p$,3$d$,4$s$ states as valence states for Mn, 3$s$,3$p$,4$s$ states for Ca and 2$s$,2$p$ as valence states for O. The wave-function was expanded in plane-waves, upto a  kinetic energy cutoff of 55 Ha. Integrals over the Brillouin zone were replaced by sums on a 6$\times$6$\times$6 mesh of special $k$-points. The phonons, dielectric tensors and Born effective charges have been calculated in the framework of density functional perturbation theory~\cite{dfpt}.

First, we reinvestigated the structural, electronic and dielectric properties of CaMnO$_3$ in the cubic perovskite structure. In this high symmetry phase, the atomic positions are fixed by symmetry so that the only structural degree of freedom is the lattice parameter. Our calculations provide a relaxed lattice constant $a_0 = 3.74$ \AA, which is in excellent agreement with the experimental estimate of 3.73 \AA \cite{woll}, illustrating the accuracy of the Wu and Cohen functional for magnetic compounds. 

\begin{table}[[htbp!]
\begin{tabular}{ccccccccc}
\hline
\hline
 &$a_{cell}$ & E$_g$ & $Z^*_{Ca}$ & $Z^*_{Mn}$ & $Z^*_{O_\bot}$ & $Z^*_{O_{\rVert}}$ &$\epsilon^{\infty}$\\
 & \AA & $eV$ & $|e|$ & $|e|$ & $|e|$ & $|e|$ & $-$ \\
\hline
\hline
$a_{-2\%}$ & 3.67&0.64  &2.63 &7.75 &-1.79 &-6.80&10.42\\
\hline
 $a_{0}$& 3.74 &0.54 &2.61 &8.16 &-1.80 &-7.18&11.49\\
\hline
$a_{+2\%}$ & 3.81 &0.41 &2.60 &8.68 &-1.82 &-7.64&13.24\\
\hline
\hline
\end{tabular}
\caption{Lattice constant, energy gap, Born effective charges and optical dielectric constant of cubic CaMnO$_3$ at three different volumes. $O_\bot$ and $O_{\rVert}$ refer respectively to O displacements parallel and perpendicular to the Mn--O direction.}
\label{tab1}
\end{table}
Within GGA, this relaxed structure is correctly predicted to be insulating.  Although significantly smaller than the experimental value (3.1 eV\cite{gap}) because of the well-known DFT bandgap problem, our calculated bandgap of 0.54 eV nevertheless agrees with previous LDA calculations \cite{fil,fil2}. On top of calculations at the optimized lattice parameter, we also considered structures at $a_{-2\%}= 0.98$ $a_0 = 3.67$ \AA $\,$ and $a_{+2\%}= 1.02$~$a_0 = 3.81$ \AA. The results are summarized in Table I. We observe that the electronic bandgap decreases when increasing the volume, which is expected from the evolution of the crystal field in a compound where the gap arises from the splitting between $t_{2g}$ and $e_g$ orbitals. CaMnO$_3$ nevertheless remains an insulator at the three volumes, even at the GGA level.

The Born effective charges (Z$^*$) and optical dielectric constants ($\epsilon_{\infty}$) are also reported in Table I and compare well with previous calculations \cite{fil,fil2}. We notice that the Born effective charges present features very similar to those of ferroelectric perovskites \cite{Ghosez-Z}, with  Z$^*_{Mn}$ and Z$^*_{O\parallel}$ (for O displacement parallel to the Mn--O direction) being {\it anomalously} large. In prototypical ferroelectric perovskites like BaTiO$_3$, these anomalous Z$^*$ were shown to be a key feature to produce a giant destabilizing dipolar interaction at the origin of the ferroelectric instability \cite{phil}. Although the dielectric constant is larger in CaMnO$_3$ as properly noticed in Ref.~\cite{fil}  and will so better screen the Coulomb interaction, it remains surprizing that CaMnO$_3$ does not exhibit any tendency to ferroelectricity in spite of its giant Z$^*$.

We therefore computed the phonon frequencies within the cubic perovskite structure in order to identify the structural instabilities inherent to CaMnO$_3$. At the optimized volume we surprisingly identified a small FE instability related to a unstable polar mode with an imaginary frequency $\omega_{FE} = 13 i$ cm$^{-1}$. However, we also observed, within the same structure, a much larger AFD instability associated to a non-polar oxygen rotational mode with an imaginary frequency $\omega_{AFD} = 213 i$ cm$^{-1}$. It is in fact the condensation of this latter mode with different amplitudes along the different directions that generates the observed orthorhombic ground-state. The case of CaMnO$_3$ is therefore very similar to that of SrTiO$_3$ or CaTiO$_3$: it develops both FE and AFD instabilities and the absence of a ferroelectric ground-state is not directly related to its magnetic character but to a much larger tendency to oxygen rotations. Both types of instability are in competition and we explicitely checked that once the dominant AFD distrortion is condensed, the small FE instability is suppressed by anharmonic effects.
\begin{center}
\begin{figure}[htbp]
{\par\centering
 {\scalebox{0.4}{\includegraphics{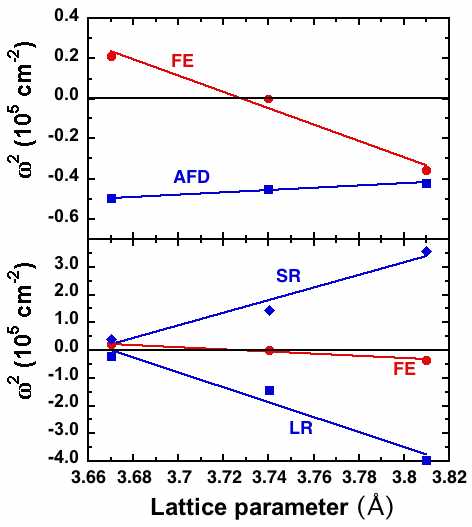}}}
 %{\scalebox{0.30}{\includegraphics{Fig2.jpg}}}
\par}
\caption{(Color online) {\bf Top: } Evolution of the square of the frequency of FE mode (red circles) and AFD mode (blue squares) of cubic CaMnO$_3$ with  the
lattice constant. {\bf Below}: Decomposition of the square of the FE frequency
(red circles) into its short-range (blue diamonds) and long-range contributions
(blue squares) as explained in the text. } 
\label{fig1}
\end{figure}
\end{center}
Performing a similar characterization at different volumes, we observe in Fig. 1 that the AFD instability is rather independent of the lattice constant. At the opposite, the FE instability is strongly sensitive to volume, a result totally similar to what was previously reported for titanates \cite{book}. This might explain why no FE instability was previously reported from LDA calculations which typically underestimates the volume by 1-2 \%.

\begin{center}
\begin{figure}[htbp]
{\par\centering
 {\scalebox{0.4}{\includegraphics{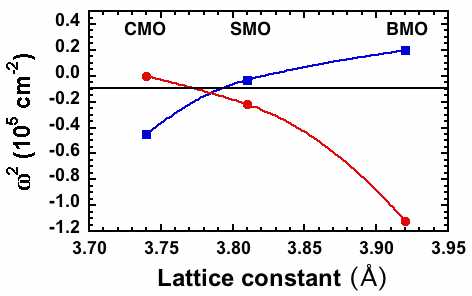}}}
\par}
\caption{(Color online) Square of the frequency of FE (red circles) and AFD (blue squares) modes of CaMnO$_3$ (CMO, $a_0 = $3.74 \AA), SrMnO$_3$ (SMO, $a_0 = $3.81 \AA) and BaMnO$_3$ (BMO, $a_0 = $3.92 \AA) at their own equilibrium volume.} 
\label{fig2}
\end{figure}
\end{center}

Following the original idea of Cochran\cite{coch}, the ferroelectric instability in perovskite oxides is usually explained in terms of a close competition between short-range forces, which tend to favor the high symmetry phase, and long-range Coulomb interactions which tend to destabilize the structure. In order to clarify wether this idea still apply to CaMnO$_3$ or wether its ferroelectric instability has a different origin, we performed a decomposition of the FE frequency in terms short-range (SR) and long-range (LR) Coulomb contribution following a scheme previously applied to BaTiO$_3$ in Ref.~\cite{phil} : 
\begin{eqnarray} \label{SR-LR_Decompo}
\underbrace{\langle \gamma_{FE} | \tilde{D}_{tot} |  \gamma_{FE} \rangle}_{\textstyle\omega^2_{FE}} 
= \underbrace{\langle  \gamma_{FE} | \tilde{D}_{LR} |  \gamma_{FE} \rangle}_{\textstyle\omega^2_{LR}}
+ \underbrace{\langle  \gamma_{FE} | \tilde{D}_{SR} |  \gamma_{FE} \rangle}_{\textstyle\omega^2_{SR}}
\nonumber
\label{Eq-1}
\end{eqnarray}
where $\tilde{D}_{tot}$ refers to the full dynamical matrix at $\Gamma$, $\tilde{D}_{LR}$ and $\tilde{D}_{SR}$ to its LR and SR contributions and \textbar$ \gamma_{FE} \rangle$ is the dynamical matrix eigenvector of the FE mode. The results are summarized in Figure~\ref{fig2} and are comparable to those previously reported for BaTiO$_3$ : the small FE instability is a consequence of a delicate balance between SR and LR forces and arises from a giant negative LR contribution directly related to the anomalously large $Z^*$ values. We explicitely checked that, considering them at their own optimized volume, CaMnO$_3$ develops  a DD contribution ($\omega_{DD} = 378.4 i $ cm$^{-1}$) even larger to that of CaTiO$_3$ ($\omega_{DD} =  345 i $ cm$^{-1}$)  and that in spite of its larger dielectric constant. The smaller FE instability of CaMnO$_3$ is related to  stronger SR forces associated to its smaller equilibrium volume ($\omega_{SR} = 378.2 $ cm$^{-1}$ in CaMnO$_3$ versus $\omega_{SR} = 313$ cm$^{-1}$ in CaTiO$_3$) .

The natural tendency of CaMnO$_3$ to develop a ferroelectric distortion being established, it might now be questioned which cation is driving the FE instability. To clarify this issue, the involvement of each type of atom within the ferroelectric soft-mode eigendisplacement vector is reported in Table II \cite{Note}. Inspection of this Table points out  that the progressive softening of the FE mode when increasing the volume is closely related to the concomitant increase of the Mn atom motion. This involvement of the Mn atom which has the largest $Z^*$, combined to a  lower extent with the small increase of $Z^*$ itself, produce a significant increase of the mode effective charge and of the related destabilizing DD interaction. This clearly establishes the Mn driven character of the ferroelectric instability. CaMnO$_3$ provides therefore a first prototypical example of magnetic perovskite ABO$_3$ compound in which ferroelectricity and magnetism are driven by the {\it same} cation, illustrating the fact that magnetism and ferroelectricity are not necessarily exclusive.

\begin{table}[htbp]
\begin{tabular}{lccccccc}
\hline
\hline
ABO$_3$ & $a_{cell}$ & $\omega_{AFD}$ &$ \omega_{FE}$ & $Z^*_{FE}$ & $2\eta^2_{A}$  &$2\eta^2_{B}$  &$6\eta^2_{O}$ \\
& \AA  & cm$^{-1}$ & cm$^{-1}$  &|e|   \\
\hline
CaMnO$_3$ & 3.67  &223i  &146   &7.51     & 0.59 & 0.00 &0.41 \\
                     & 3.74       &213i  &13i   & 10.22  & 0.34 &0.05 &0.61  \\
                     & 3.81 &205i  &189i  & 13.48  & 0.08 & 0.26 &0.66    \\
SrMnO$_3$ & 3.81  &55i  &148i   &14.65     & 0.00 & 0.19 &0.81 \\
Ca$_{0.5}$Ba$_{0.5}$MnO$_3$ & 3.83  &150i  &216i   &15.28     & 0.00 & 0.21 &0.79 \\
BaMnO$_3$ & 3.92  &140  &335i   &19.24     & 0.00 & 0.22 &0.78 \\
\hline
\hline
\end{tabular}
\caption{Lattice constant ($a_{cell}$), frequencies of the AFD ($\omega_{AFD}$) and FE ($\omega_{FE}$) phonon modes, mode effective charge of the FE mode ($Z^*_{FE}$) and involvement of each type of atom to the FE mode eigendisplacement vector \cite{Note}  for different compounds in the cubic perovskite structure. For Ca$_{0.5}$Ba$_{0.5}$MnO$_3$, results correspond to an ordered supercell in witch Ca and Ba alternate along the [111] direction.}
\label{tab2}
\end{table}
The absence of FE phase transition in CaMnO$_3$ being related to the competition with the AFD instability, we now investigate to which extent it would be possible to favor ferroelectricity and so, eventually, make CaMnO$_3$ multiferroic using either strain or chemical engineering. 

From Fig.1, it appears that FE and AFD instabilities exhibit a very different sensitivity to strain, a fact that can be potentially exploited to  favor the ferroelectric distortion. Since applying a negative isotropic pressure is not easily achievable in practice, we rather explore the possibility to make CaMnO$_3$ ferroelectric under epitaxial strain as it could be achieved  in epitaxial thin films. In ferroelectric oxides, it is well-known \cite{Dieguez} that large enough compressive epitaxial strains typically favor a ferroelectric c-phase (with the polarization out-of-plane)  while tensile strains favor a ferroelectric a-phase (with polarization in-plane) but no systematic data are available for the impact of epitaxial strain on the AFD distortion.  Exploring first the effect of compressive epitaxial strains, we observed that the ground-state of CaMnO$_3$ remains AFD distorted and unfortunately paraelectric for strains as large as $-4$\%, roughly the largest epitaxial strain experimentally achievable. At the opposite, for tensile strains, it appeared however that a polar ground-state is obtained for strains larger than about $+2$\%. Beyond this critical value, the system preserves its oxygen rotations by aquires an additional in-plane polarization estimated from a Berry phase calculation to 4 $\mu$C.cm$^{-2}$ at an in-plane lattice constant of 3.82 \AA. This offers the theoretical demonstration that CaMnO$_3$ can be made ferroelectric by strain engineering.

Another possible way to increase the volume and so provide more space for the FE distortion of the Mn atom is to perform atomic substitution and to replace Ca by a larger ion. To support this idea we compare in Figure  2 and Table II the behavior of CaMnO$_3$, SrMnO$_3$ and BaMnO$_3$,  each of them in the same prototypical cubic perovskite structure at their own equilibrium volume.  It clearly appears that increasing the size of the A cation strongly favor the FE distortion that is the only remaining instability in BaMnO$_3$. Going further we see that the atomic substitution goes beyond a simple volume effect since together with favoring ferroelectricity, bigger A cations also suppress the possibility of oxygen rotation. From that viewpoint, atomic substitution of Ca by a bigger atom appears even more promizing than epitaxial strain to produce a ferroelectric ground-state. 

To support this idea, we performed calculations on alloy Ca$_{0.5}$Ba$_{0.5}$MnO$_{3}$. Working first within the virtual crystal approximation (VCA) \cite{BST}, we obtained a relaxed lattice constant $a_0= 3.84$ \AA $\,$ and observed a strong FE instability ($\omega_{FE} =  224 i $ cm$^{-1}$) while the AFD instability was totally suppressed ($\omega_{AFD} =  110 $ cm$^{-1}$). Considering then an ordered supercell in which Ca and Ba ions alternates along the [111] direction,  the relaxed lattice constant ($a_0 = 3.83$ \AA) and the ferroelectric instability ($\omega_{FE} =  215 i $ cm$^{-1}$) were very similar to VCA but the AFD mode was now unstable ($\omega_{AFD} =  150 i $ cm$^{-1}$). This points out that the amplitude of the AFD mode might be very sensitive to the atomic order but both approaches predict a dominant ferroelectric instability. This confirms that, eventually combined with strain engineering, partial substitution of Ca by Ba will favor ferroelectricity and might be also a promizing way to make CaMnO$_3$ multiferroic. 
In conclusion, we have investigated from first-principles the lattice dynamics of cubic CaMnO$_3$. We pointed out that, together with a strong AFD instability responsible for its orthorhombic ground-state, CaMnO$_3$ also surprisingly exhibits a weak FE instability at its equilibrium volume. Going further, we highlighted that both strain and chemical engineerings can be used to favor the ferroelectric distortion and make  CaMnO$_3$ multiferroic. Finally, we also showed that the FE instability of CaMnO$_3$ is Mn-dominated, illustrating that ferroelectricity and magnetism are not necessarily exclusive and can be driven by the same cation. All this suggests alternative strategies in the search of multiferroic ABO$_3$ compounds and we hope that our findings will motivate further experimental investigations.

This work was supported by the European STREP MaCoMuFi, the European FAME-EMMI, NoE. The simulations have been performed on the supercomputer MareNostrum at  the Barcelona Supercomputing Center - Centro Nacional de Supercomputacion.


\begin{thebibliography}{99}
\bibitem{Note}
We separated the contribution of each atom to the eigendisplacement vector $\eta$ which is related to the dynamical matrix eigenvector $\gamma$ through : $|\eta> = |\gamma>/\sqrt{M}$.
\end{thebibliography}
\end{document}